\documentclass[conference]{IEEEtran}
\IEEEoverridecommandlockouts
\def\BibTeX{{\rm B\kern-.05em{\sc i\kern-.025em b}\kern-.08em
    T\kern-.1667em\lower.7ex\hbox{E}\kern-.125emX}}

\usepackage{indentfirst}
\usepackage{graphicx}
\usepackage{amssymb}
\usepackage{amsfonts}
\usepackage{mathrsfs}
\usepackage{leftidx}
\usepackage{color}
\usepackage{amsmath}
\usepackage{arydshln}
\usepackage{ragged2e}
\usepackage{cite}
\usepackage{enumerate}
\usepackage{longtable}
\usepackage{float}
\usepackage{stfloats}
\usepackage{algpseudocode}
\usepackage{algorithm}
\usepackage[caption=false,font=normalsize,labelfont=sf,textfont=sf]{subfig}
\usepackage{tabularx}

\usepackage[table,usenames,dvipsnames]{xcolor}
\definecolor{aa}{RGB}{175,238,238}
\definecolor{bb}{RGB}{255,255,255}

\usepackage{bm}
\usepackage{makecell}

\begin{document}

\title{KGRAG-SC: Knowledge Graph RAG-Assisted Semantic Communication}

\author{
\IEEEauthorblockN{Dayu Fan, Rui Meng, Song Gao, Xiaodong Xu}
\IEEEauthorblockA{ State Key Laboratory of Networking and Switching Technology, BUPT, Beijing, China}

\IEEEauthorblockA{\{fandayu, buptmengrui, wkd251292, xuxiaodong\}@bupt.edu.cn}
\thanks{This work was supported in part by the National Key Research and Development Program of China under Grant 2020YFB1806905; in part by the National Natural Science Foundation of China under Grant 62501066 and under Grant U24B20131; and in part by the Beijing Municipal Natural Science Foundation under Grant L242012. \textit{(Corresponding author: Rui Meng)}}
}

\maketitle

\begin{abstract}
The state-of-the-art semantic communication (SC) schemes typically rely on end-to-end deep learning frameworks that lack interpretability and struggle with robust semantic selection and reconstruction under noisy conditions. To address this issue, this paper presents KGRAG-SC, a knowledge graph-assisted SC framework that leverages retrieval-augmented generation principles. KGRAG-SC employs a multi-dimensional knowledge graph, enabling efficient semantic extraction through community-guided entity linking and GraphRAG-assisted processing. The transmitter constructs minimal connected subgraphs that capture essential semantic relationships and transmits only compact entity indices rather than full text or semantic triples. An importance-aware adaptive transmission strategy provides unequal error protection based on structural centrality metrics, prioritizing critical semantic elements under adverse channel conditions. At the receiver, large language models perform knowledge-driven text reconstruction using the shared knowledge graph as structured context, ensuring robust semantic recovery even with partial information loss. Experimental results demonstrate that KGRAG-SC achieves superior semantic fidelity in low Signal-to-Noise Ratio (SNR) conditions while significantly reducing transmission overhead compared to traditional communication methods, highlighting the effectiveness of integrating structured knowledge representation with generative language models for SC systems.

\end{abstract}

\section{Introduction}

Oriented towards 6th Generation application scenarios such as autonomous driving and human-machine symbiotic intelligence, communication systems are facing unprecedented challenges with explosive growth in data transmission demands and limited bandwidth availability \cite{fan2025generative,wu2025lotterycodec,cao2025importance}. Semantic Communication (SC) has emerged to meet this demand, jointly optimizing the source and channel to preserve meaning, demonstrating significant potential under limited bandwidth and low Signal-to-Noise Ratio (SNR) conditions\cite{lu2025important,meng2025survey,wu2025actions}. While deep learning-based end-to-end frameworks have shown promise, they often function as ``black boxes", facing critical challenges in interpretability, robustness, and generalization, especially when dealing with the vast and rapidly evolving body of real-world knowledge \cite{10854543,10963886}.

A natural and powerful solution is to integrate external, structured knowledge bases into the communication process. Knowledge Graphs (KGs), which represent information as explicit ``entity-relation-entity" triples, are particularly well-suited for this role. By grounding communication in a shared, interpretable KG, it becomes possible to align semantics, perform robust inference, and achieve meaningful compression. Early efforts to integrate KGs into SC have validated this ``knowledge-assisted" approach, demonstrating improved performance\cite{wang2023knowledge}. However, these works also revealed practical limitations. 
Firstly, due to the introduction of unnecessary overhead caused by sending redundant or excessively large graphic structures, it is necessary to quickly and accurately select the most significant semantic information from KGs for transmission. Furthermore, these schemes often lack a sophisticated mechanism to fully leverage the rich context within the KG for high-fidelity text reconstruction at the receiver.

In parallel, the field of natural language processing has seen the rise of Retrieval-Augmented Generation (RAG)\cite{lewis2020retrieval}, a powerful technique that enhances the factuality and relevance of Large Language Models (LLMs) by conditioning their outputs on externally retrieved information. More advanced paradigms like GraphRAG\cite{edge2024local,han2024retrieval} extend this concept to structured knowledge, retrieving connected subgraphs to provide LLMs with richer, relational context that is crucial for complex reasoning. This principle of ``retrieve-then-generate" over a structured knowledge base offers an effective approach for selecting salient, context-aware information, directly addressing the core challenges faced in SC schemes.

Inspired by these advancements, we propose a novel Knowledge Graph-assisted SC framework, which we refer to as KGRAG-SC (Knowledge Graph RAG-assisted Semantic Communication), that adapts the GraphRAG philosophy to the communication pipeline. At the transmitter, we perform a context-aware retrieval to identify a Minimum Connected Subgraph (MCSG) that preserves the core meaning of the source text, and then transmit only its compact index representation. At the receiver, this structured subgraph, along with rich entity descriptions from the KG, serves as a grounded prompt for an LLM to perform high-fidelity text reconstruction. Experimental results demonstrate that KGRAG-SC achieves superior semantic fidelity with significantly reduced transmission overhead compared to traditional methods. This design achieves an effective balance between compression efficiency, interpretability, and robustness. The main contributions of this paper are as follows:

\begin{itemize}
    \item \textbf{Multi-dimensional KG and efficient retrieval representation:} A local Knowledge Graph with ``community stratification + node description" is constructed offline based on WebNLG. The ``all-MiniLM-L6-v2" model is used to generate 384-dimensional vectors, and FAISS is adopted to establish indexes. Meanwhile, an ``entity to ID" mapping is maintained, laying a foundation for transmitting only IDs in subsequent steps.
    \item \textbf{Minimum Connected Subgraph and extreme compression transmission:} Guided by GraphRAG, normalized entity relations are stably extracted, and a MCSG that maintains semantic coherence is constructed. Only the list of node IDs constituting the MCSG is transmitted instead of the original triples, realizing high compression ratio and low redundancy.
    \item \textbf{``KG + LLM" semantic reconstruction and adaptive robustness:} At the receiver, the shared KG and LLM are used to restore the subgraph to natural language; at the physical layer, adaptive channel coding (16QAM + Convolutional Code) is performed based on node importance to prioritize the protection of key nodes, significantly improving the recovery performance under low SNR.
\end{itemize}

\section{Related Work}
This section provides a review of the existing literature, focusing on two key areas: the integration of knowledge graphs into SC, and the application of RAG techniques that inspire our approach.

\subsection{Knowledge Graph-Aided SC}
The integration of knowledge graphs into SC has emerged as a promising approach to address the fundamental challenge of semantic representation. Knowledge graphs, with their structured ``entity-relation-entity" triplet format, provide an interpretable and efficient means to encode semantic information while reducing transmission overhead.

Early foundational work by Jiang et al. \cite{jiang2022reliable} established the basic paradigm of knowledge graph-enabled SC. In their schemes, transmitted sentences are converted into semantic triplets using deep learning-based extraction methods, where triplets serve as basic semantic symbols that can be sorted according to semantic importance. The scheme adaptively adjusts transmitted content based on channel quality, allocating more transmission resources to important triplets through unequal error protection schemes. This work demonstrated significant improvements in communication reliability under low signal-to-noise ratio conditions compared to traditional schemes.

Building upon this foundation, Ren et al. \cite{ren2024knowledge} proposed a more comprehensive knowledge base framework from a generative perspective. Their approach divides the semantic knowledge base into three sub-components: source KB, task KB, and channel KB, each addressing different aspects of the communication process. The source KB serves as the core generative component, while task and channel KBs provide contextual information to guide semantic representation. This multi-faceted approach aims to overcome the limitations of single-modality, single-task scheme by enabling cross-modal fusion and cross-environment transmission.

Recent advances have focused on optimizing the semantic extraction and transmission strategies within knowledge graph-aided scheme. Wang et al. \cite{wang2022performance} proposed an attention-based reinforcement learning approach for performance optimization in SC. Their work specifically addresses the challenge of selective transmission by modeling semantic information as knowledge graphs consisting of semantic triples, and using an attention mechanism to evaluate the importance of each triple for optimal resource allocation and partial semantic information transmission. This attention-based mechanism provides a conceptual framework for the importance-aware transmission strategies explored in our work.

However, existing knowledge graph-aided schemes face several critical limitations that our work addresses. First, current approaches typically rely on simple entity extraction from knowledge graphs and basic graph construction methods, lacking rapid and accurate matching methods for large-scale knowledge graphs in context-aware semantic extraction and compression. Second, most schemes transmit either full semantic triplets or basic entity indices without optimizing for minimal connected subgraph structures that preserve semantic coherence, resulting in redundant transmission content. Third, the reconstruction process at the receiver often involves simplistic template-based or rule-based text generation, failing to leverage the rich contextual information available in knowledge graphs for high-fidelity recovery. KGRAG-SC addresses these gaps by incorporating RAG principles for robust semantic extraction, constructing minimal connected subgraphs for extreme compression, and utilizing large language models with structured knowledge prompts for superior text reconstruction.

\subsection{RAG and Structured Knowledge}
RAG\cite{lewis2020retrieval} has recently become a mainstream standard for mitigating hallucination and incorporating external, up-to-date knowledge into LLM outputs. The core idea of RAG is to first retrieve relevant documents or data snippets from a large corpus and then use this retrieved information as context for the LLM during the generation phase.

While standard RAG typically operates on unstructured text corpora, there is a growing interest in applying similar principles to structured knowledge bases like KGs. This has led to the development of GraphRAG \cite{edge2024local}, a technique that retrieves information from graph-structured data to augment LLM prompts. Instead of retrieving flat text chunks, GraphRAG can traverse the graph to find interconnected entities and relationships, providing the LLM with a richer, more structured context. This allows the model to perform more complex reasoning and generate responses that are grounded in the factual relationships defined within the KG.

KGRAG-SC is conceptually aligned with the GraphRAG philosophy. At the transmitter, we perform a form of knowledge retrieval and consolidation by identifying entities and mapping them to a minimal subgraph in the KG. This structured, compressed representation is then transmitted. At the receiver, the reconstruction process provides the LLM with structured evidence from the KG, specifically the subgraph topology and corresponding entity descriptions, guiding it to generate a faithful textual representation. By adapting this retrieval-augmented mindset to the domain of SC, we aim to build a scheme that is not only efficient but also more robust and interpretable.

\section{Proposed Scheme: KGRAG-SC}
This section introduces the overall architecture of KGRAG-SC, and elaborates on the design principles and implementation methods of each key module in sequence.

\begin{figure}[htbp]
\centering
\includegraphics[width=\linewidth]{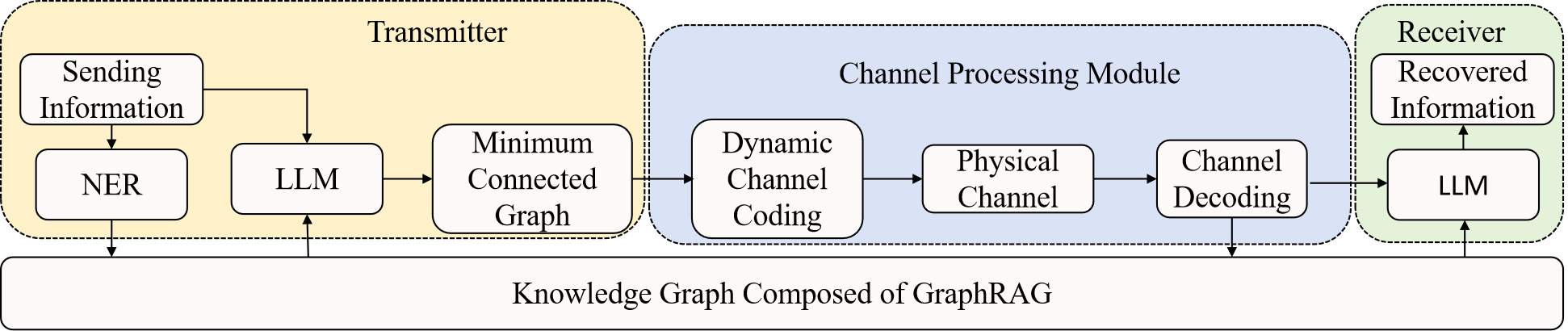}
\caption{The overall architecture of KGRAG-SC, detailing the workflow from semantic extraction at the transmitter to text reconstruction at the receiver.}
\label{fig1}
\end{figure}

\subsection{Scheme Overview}
The end to end workflow of KGRAG-SC is depicted in Fig.~\ref{fig1}. The framework is composed of three primary components: a transmitter, a physical channel, and a receiver. All components operate on a shared, preconstructed KG.

\textbf{Transmitter:} The core task at the transmitter is to transform unstructured source text into a highly compressed, structured semantic representation. This process begins with semantic extraction and linking, where KGRAG-SC identifies key entities within the source text and links them to their corresponding nodes in the shared KG, thereby grounding the text in a canonical knowledge base. Next, for semantic compression, we construct a MCSG that encapsulates the core relationships between the linked entities, rather than transmitting raw text or full semantic triples. This graph is then distilled into a sequence of node indices, achieving a high compression ratio. Finally, KGRAG-SC employs importance aware channel coding. Recognizing that not all semantic nodes are equally important, we calculate an importance score for each node in the MCSG. This score guides an adaptive channel coding scheme that provides stronger error protection to more critical information before transmission.

\textbf{Channel:} The encoded bitstream is then modulated and transmitted over the physical channel, where it is inevitably corrupted by noise.

\textbf{Receiver:} At the receiver, the goal is to robustly reconstruct the original meaning from the potentially erroneous received signal. The process starts with subgraph reconstruction. After demodulation and channel decoding, the received node indices are used to look up the corresponding entities in the shared KG, thereby reconstructing the MCSG. Following this, KGRAG-SC performs text generation driven by knowledge. The reconstructed subgraph, along with rich contextual information such as entity descriptions retrieved from the KG, is formatted into a structured prompt. This prompt is then fed to a LLM, which performs constrained generation to produce the final, human readable text. This architecture synergistically combines the structured, factual grounding of KGs with the powerful inferential and generative capabilities of LLMs to create a communication scheme that is both efficient and robust.

\subsection{Transmitter Design: From Text to Structured Semantics}
The transmitter's design focuses on efficiently encoding textual information into a minimal set of structured indices for transmission.

\subsubsection{Foundational Knowledge Base: The Multi-dimensional
Knowledge Graph}
The shared KG serves as the backbone of KGRAG-SC. We construct it offline from the WebNLG dataset, which is rich in triples of entity, relation, and entity. The KG is formally defined as a tuple $G=(V, E, \mathcal{D}, \mathcal{C})$. Here, $V$ represents the set of all entity nodes. $E$ is the set of directed edges, which are defined as $V \times R \times V$ where $R$ is the set of relation types. The function $\mathcal{D}: V \to T$ maps each entity to its textual description $T$. Lastly, $\mathcal{C}: V \to C$ maps each entity to a community or category $C$, such as ``University" or ``City", which provides a hierarchical structure. This three dimensional structure, encompassing nodes, descriptions, and communities, provides a rich, multifaceted prior for both semantic extraction and reconstruction. To enable efficient lookup, all entity names and descriptions are embedded into a 384 dimensional vector space using the \texttt{all-MiniLM-L6-v2} model\cite{reimers2019sentence,wang2020minilm}. This model is chosen for its balance of performance and computational efficiency. The resulting vectors are indexed using FAISS\cite{johnson2019billion} (Facebook AI Similarity Search) to facilitate rapid, large scale similarity searches during the entity linking phase.

\subsubsection{GraphRAG-Assisted Entity Extraction}

Given an input sentence $S$, our goal is to extract the most relevant entities that accurately represent its semantic content. This is achieved through a three-stage GraphRAG-assisted process that leverages both neural entity recognition and knowledge graph retrieval to guide LLM-based entity selection.

\textbf{Stage 1: Initial Entity Recognition.} We first employ a standard Named Entity Recognition (NER) model to identify a preliminary set of candidate entities $E_{\text{ner}}$ from the input text:
\begin{equation}
    E_{\text{ner}} = f_{\text{NER}}(S)
\end{equation}

\textbf{Stage 2: Community-Guided Candidate Expansion.} Following the GraphRAG paradigm, we employ a two-step hierarchical matching strategy to efficiently identify relevant entities while minimizing computational overhead. For each NER-identified entity $e \in E_{\text{ner}}$, we first match it against community summaries to identify the most relevant community:
\begin{equation}
    c_e^* = \arg\max_{c \in \mathcal{C}} \frac{\mathbf{v}_e \cdot \mathbf{s}_c}{\|\mathbf{v}_e\| \|\mathbf{s}_c\|}
\end{equation}
where $\mathbf{s}_c$ is the precomputed embedding of community $c$'s summary. Once the target community is identified, we perform fine-grained similarity search only within that community to find the top-3 most relevant entities:
\begin{equation}
    C_e = \text{Top3}_{v \in V_{c_e^*}} \left( \frac{\mathbf{v}_e \cdot \mathbf{v}_v}{\|\mathbf{v}_e\| \|\mathbf{v}_v\|} \right)
\end{equation}
This hierarchical approach dramatically reduces the search space from $|V|$ to $|V_{c_e^*}|$, where $|V_{c_e^*}| \ll |V|$, leading to significant computational savings. The union of all candidate sets forms our expanded candidate pool: $E_{\text{candidates}} = \bigcup_{e \in E_{\text{ner}}} C_e$.

\textbf{Stage 3: LLM-guided Entity Selection.} Rather than directly selecting entities, we construct a structured prompt that includes both the original sentence and the candidate entities with their KG descriptions. The LLM is then tasked with selecting the most contextually relevant entities:
\begin{equation}
    E_{\text{selected}} = f_{\text{LLM}}(S, \{(e, \text{desc}(e)) | e \in E_{\text{candidates}}\})
\end{equation}
where $\text{desc}(e)$ retrieves the descriptive information of entity $e$ from the knowledge graph.

This GraphRAG-assisted approach is formalized in Algorithm~\ref{alg:entity_extraction}, which demonstrates how the preconstructed knowledge graph guides the entity extraction process through semantic similarity and contextual relevance.

\begin{algorithm}
\caption{Community-Guided GraphRAG Entity Extraction}\label{alg:entity_extraction}
\begin{algorithmic}[1]
\Require Input sentence $S$, KG $G=(V, E, \mathcal{D}, \mathcal{C})$, Community summaries $\{\mathbf{s}_c\}$
\Ensure Selected entities $E_{\text{selected}}$

\State $E_{\text{ner}} \gets f_{\text{NER}}(S)$
\Statex \(\triangleright\) Stage 1: NER extraction
\State $E_{\text{candidates}} \gets \emptyset$

\For{each entity $e \in E_{\text{ner}}$}
    \Statex \(\triangleright\) Stage 2: Community-guided expansion
    \State $\mathbf{v}_e \gets \text{GetEmbedding}(e)$
    \State $c_e^* \gets \arg\max_{c \in \mathcal{C}} \text{CosineSim}(\mathbf{v}_e, \mathbf{s}_c)$
    \Statex \(\triangleright\) Find best community
    \State $C_e \gets \text{Top3\_InCommunity}(c_e^*, \mathbf{v}_e)$
    \Statex \(\triangleright\) Search top-3 within community
    \State $E_{\text{candidates}} \gets E_{\text{candidates}} \cup C_e$
\EndFor

\State $\text{prompt} \gets \text{ConstructPrompt}(S, E_{\text{candidates}})$
\Statex \(\triangleright\) Stage 3: LLM selection
\State $E_{\text{selected}} \gets f_{\text{LLM}}(\text{prompt})$
\State \Return $E_{\text{selected}}$
\end{algorithmic}
\end{algorithm}

\subsubsection{Semantic Compression via Minimum Connected Subgraph}
Once the relevant entities are selected through the GraphRAG-assisted process, we construct a MCSG to represent the semantic relationships between these entities in the most compact form possible. This step is crucial for achieving efficient semantic compression while preserving the essential structural information needed for accurate reconstruction.

The MCSG construction addresses a fundamental challenge in SC: how to transmit the relationships between selected entities without losing critical semantic connections. Simply transmitting the entities as an unordered set would discard valuable relational information encoded in the knowledge graph. 

KGRAG-SC focuses on capturing only the most essential connections by including nodes that are directly connected (one-hop) to the selected entities $E_{\text{selected}}$. This strategy ensures minimal transmission overhead while maintaining semantic coherence:
\begin{equation}
    G_{\text{mcsg}} = \bigcup_{e \in E_{\text{selected}}} \{v \in V : (e,v) \in E \text{ or } (v,e) \in E\} \cup E_{\text{selected}}
\end{equation}

The construction process identifies all nodes that are directly adjacent to any selected entity in the knowledge graph. This one-hop expansion captures immediate semantic relationships without introducing unnecessary intermediate nodes that might exist in longer paths. The resulting subgraph may consist of multiple disconnected components, each representing a local semantic cluster around the selected entities.

This approach prioritizes compression efficiency over global connectivity. Rather than forcing all entities into a single connected component through potentially long paths, we preserve only the most direct and semantically meaningful connections. The compression efficiency stems from the shared nature of the knowledge graph - both transmitter and receiver have access to the same graph structure, allowing reconstruction of the full semantic context from the compact node ID sequence.

The final transmission payload consists of the ordered sequence of unique node IDs from the MCSG: $L_{\text{transmission}} = \text{NodeIDs}(V_{\text{mcsg}})$. This compact representation enables the receiver to reconstruct the semantic context by retrieving the corresponding one-hop subgraphs around each transmitted entity from their local knowledge graph copy.

\subsection{Physical Layer: Importance-Aware Adaptive Transmission}
To maximize semantic fidelity under adverse channel conditions, we introduce an adaptive transmission strategy that prioritizes the most crucial semantic elements.

\subsubsection{Quantifying Semantic Importance}
The structural properties of a node within the MCSG can serve as a proxy for its semantic importance. We use a weighted combination of two standard centrality metrics\cite{barthelemy2004betweenness}: degree centrality $C_D(v)$, which captures local influence, and betweenness centrality $C_B(v)$, which measures a node's role in connecting other nodes. The betweenness centrality is defined as:
\begin{equation}
    C_B(v) = \sum_{s \neq v \neq t \in V_{\text{mcsg}}} \frac{\sigma_{st}(v)}{\sigma_{st}}
\end{equation}
where $\sigma_{st}$ is the total number of shortest paths from node $s$ to $t$, and $\sigma_{st}(v)$ is the number of those paths passing through $v$. Before combining them, we normalize each score to the range [0, 1] using min max scaling. The final importance score $I(v)$ is:
\begin{equation}
    I(v) = \alpha \cdot \hat{C}_D(v) + (1-\alpha) \cdot \hat{C}_B(v)
\end{equation}
where $\hat{C}_D$ and $\hat{C}_B$ are the normalized scores and $\alpha \in [0,1]$ is a hyperparameter balancing the two metrics.

\subsubsection{Unequal Error Protection Scheme}
Based on the importance score $I(v)$, we implement an Unequal Error Protection scheme. The node IDs are partitioned into importance classes based on predefined thresholds. High importance nodes are encoded using rate-1/2 convolutional codes to provide strong error protection, while low importance nodes are transmitted without channel coding to maximize transmission efficiency. All data is modulated using 16QAM. This ensures that even if channel conditions degrade and some less critical information is lost, the core structural nodes of the MCSG are likely to be received correctly, preserving the fundamental meaning.

\subsection{Receiver Design: Knowledge-Driven Text Reconstruction}
The receiver's primary function is to reconstruct the original text by leveraging the shared KG and the generative power of an LLM.

\subsubsection{Subgraph Reconstruction and Error Handling}
Upon receiving the demodulated and decoded bitstream, the receiver obtains an estimated list of node IDs, $\hat{L}_{\text{ids}}$. It then attempts to reconstruct the MCSG by retrieving the corresponding nodes and their interconnecting edges from its local copy of the KG. This step inherently includes a form of error handling. If a received ID does not correspond to any node in the KG, a likely outcome of a bit error, it is discarded. The receiver then works with the largest connected component of the validly received nodes to form the reconstructed subgraph $\hat{G}_{\text{mcsg}}$.

\subsubsection{Constrained Text Generation with LLM}
Finally, the reconstructed subgraph $\hat{G}_{\text{mcsg}}$ is translated into a natural language prompt for the LLM. This is not a simple serialization. The prompt is carefully structured to include a list of the reconstructed triples from $\hat{G}_{\text{mcsg}}$, the textual descriptions for each node in the subgraph, and an explicit instruction to generate a coherent, natural sounding sentence that faithfully represents the relationships and entities provided. This process, modeled below, uses the rich, structured knowledge as a strong constraint on the LLM's generative process:
\begin{equation}
    \hat{S} = f_{\text{LLM}}(\text{Prompt}(\hat{G}_{\text{mcsg}}, \{\mathcal{D}(v)\}_{v \in \hat{V}_{\text{mcsg}}}))
\end{equation}
This text generation, which is driven by knowledge, prevents the LLM from ``hallucinating" facts and ensures the output is grounded in the transmitted semantic structure. It allows the schemes to recover a fluent and accurate sentence even if parts of the MCSG were lost during transmission, as the LLM can intelligently infer missing links based on the provided context.

\section{EXPERIMENTAL RESULTS AND ANALYSIS}

\subsection{Experimental Setup}
\textbf{Dataset and KG:} Our experiments are based on the WebNLG dataset\cite{gardent2017creating}. We construct a local knowledge graph containing approximately 8,000 unique triples encompassing over 300 relation types. The KG is built offline by using a large language model to process the original text associated with each triple, generating a concise description for each entity and classifying it into one of several predefined communities. This enriched information forms the backbone of our shared knowledge base.

\textbf{Models:} We utilize the \texttt{all-MiniLM-L6-v2} model to generate 384-dimensional embeddings for entities and descriptions, with FAISS enabling efficient similarity search. For all generative tasks, we employ the \texttt{Llama3.1-8B} model\cite{dubey2024llama}, a powerful open-source large language model from Meta, selected for its strong reasoning and natural language generation capabilities.

\textbf{Channel Simulation:} We simulate a physical communication link using an Additive White Gaussian Noise channel. The transmitted data is modulated using 16QAM. For KGRAG-SC's importance-aware scheme, an unequal error protection strategy is employed: based on an SNR-dependent importance threshold, nodes deemed highly important are protected by a rate-1/2 convolutional code with constraint length 7, while less important nodes are transmitted without channel coding to improve efficiency.

\textbf{Baselines:} To evaluate semantic recovery, our baseline ``traditional communication scheme" involves transmitting the source text using Huffman coding for compression followed by 16QAM modulation, without any channel coding. For transmission efficiency, we compare KGRAG-SC against raw ASCII and Huffman-coded text transmission.

\textbf{Evaluation Metric:} The primary metric for performance is semantic similarity, calculated as the cosine similarity between the sentence embeddings of the original and reconstructed texts\cite{reimers2019sentence}. These embeddings are generated using the same \texttt{all-MiniLM-L6-v2} model.

\subsection{Semantic Recovery Performance}
\begin{figure}[htbp]
\centering
\includegraphics[width=0.9\linewidth]{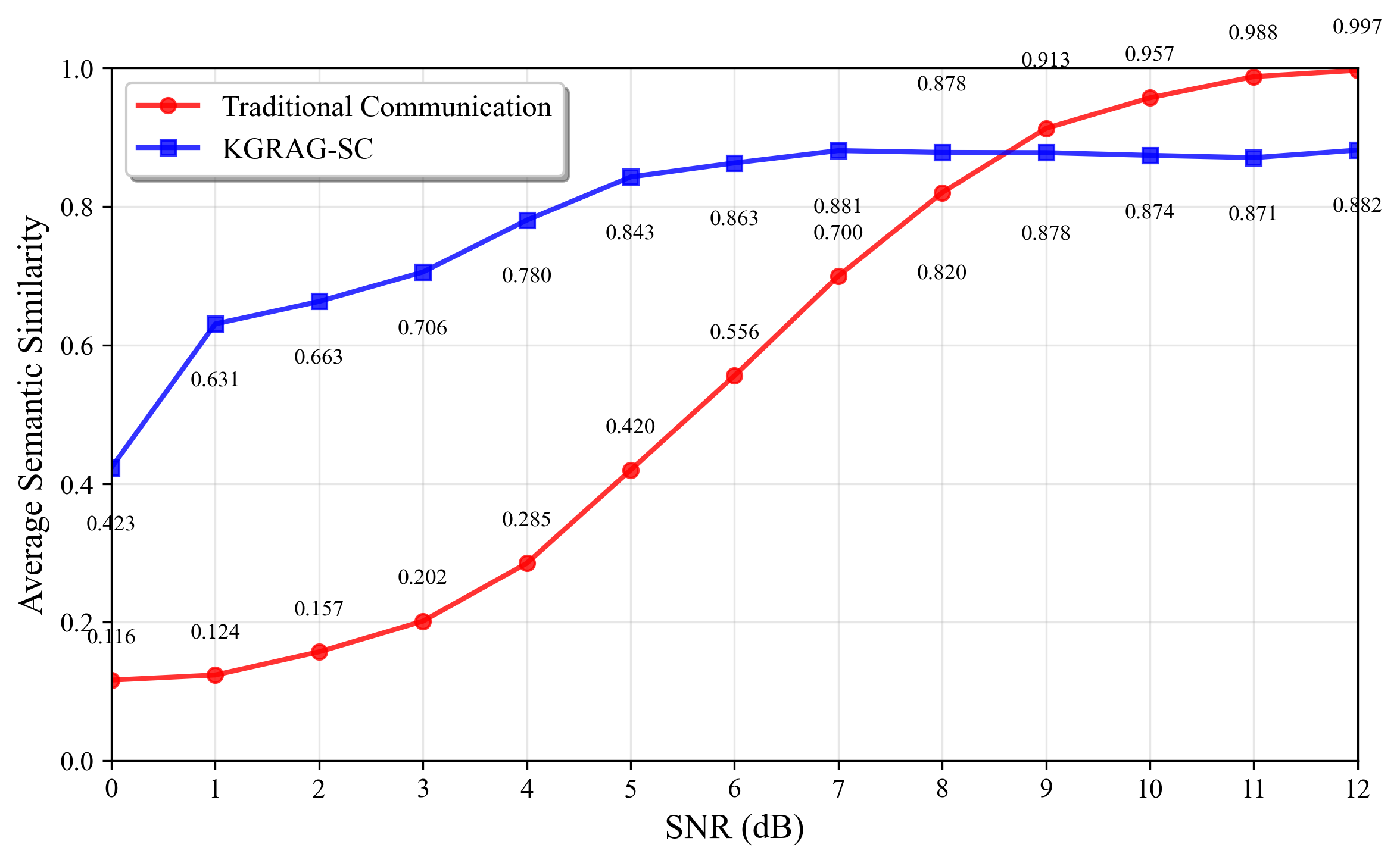}
\caption{Comparison of average semantic similarity versus SNR for KGRAG-SC and a traditional communication scheme.}
\label{fig:sem_sim}
\end{figure}

As shown in Fig.~\ref{fig:sem_sim}, we compare the semantic fidelity of KGRAG-SC with a traditional baseline under varying SNR conditions. The results demonstrate that KGRAG-SC significantly outperforms the baseline in the challenging low-to-medium SNR region from 0-8 dB, with particularly notable gains in harsh channel conditions. For example, at an SNR of 4 dB, KGRAG-SC achieves a semantic similarity score of 0.780, whereas the traditional scheme's performance is severely degraded, reaching only 0.285. However, in high SNR conditions above 8 dB, the traditional communication method achieves superior performance, reaching near-perfect semantic fidelity of 0.997 at 12 dB compared to KGRAG-SC's 0.882.

This superior robustness in the low-SNR region validates the effectiveness of KGRAG-SC's design, which preserves the core semantic skeleton even under noisy conditions. The performance crossover at approximately 8 dB SNR reveals an interesting trade-off: while KGRAG-SC excels in harsh channel conditions where traditional methods fail, the traditional approach achieves better performance in high-quality channel conditions where bit-level accuracy becomes feasible. This suggests that KGRAG-SC is particularly valuable for bandwidth-constrained or noisy communication scenarios.

\subsection{Transmission Efficiency Analysis}
\begin{figure}[htbp]
\centering
\includegraphics[width=0.9\linewidth]{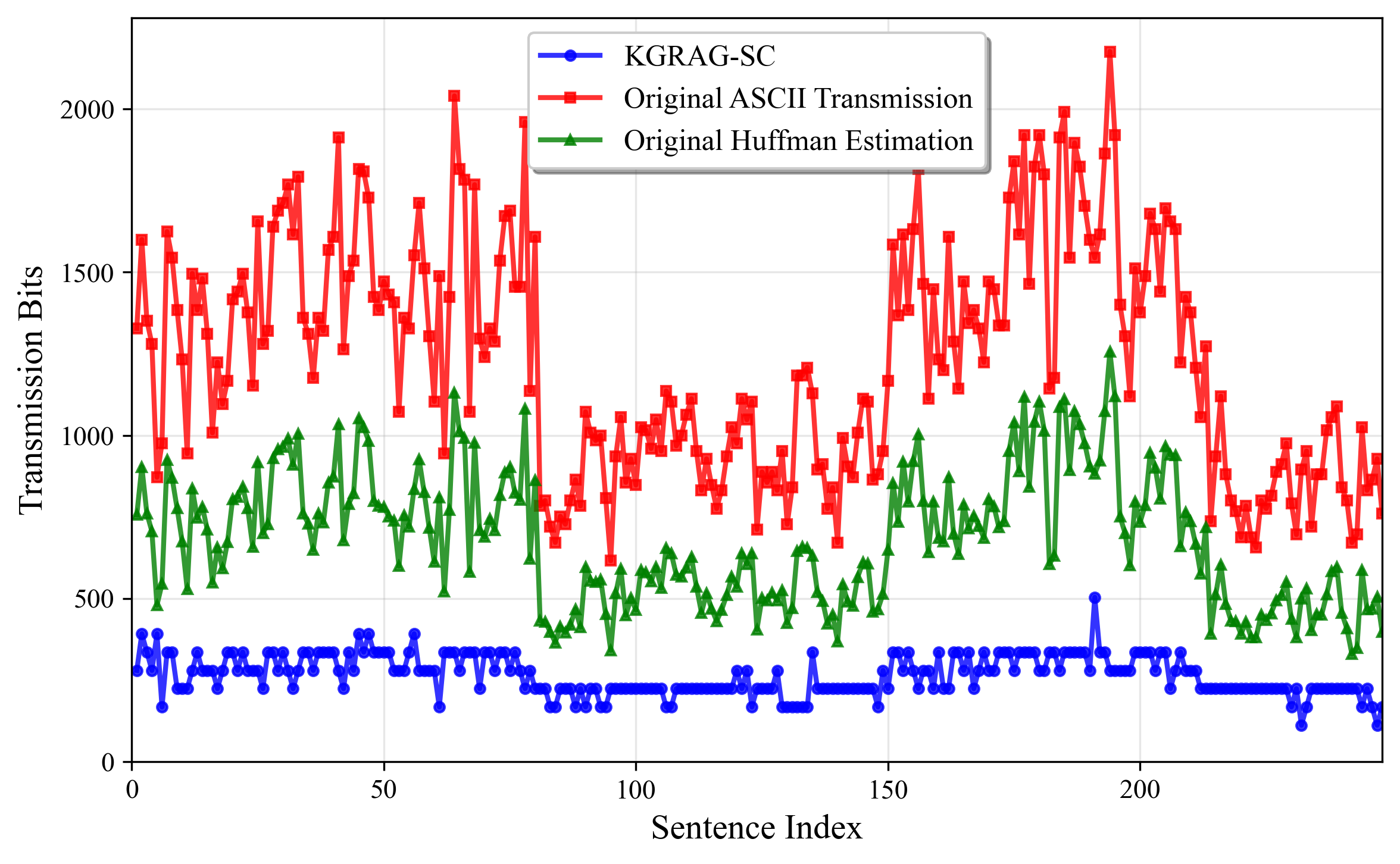}
\caption{Comparison of transmitted bits for each sentence across different transmission schemes.}
\label{fig:bits_comp}
\end{figure}

\begin{figure}[htbp]
\centering
\includegraphics[width=0.9\linewidth]{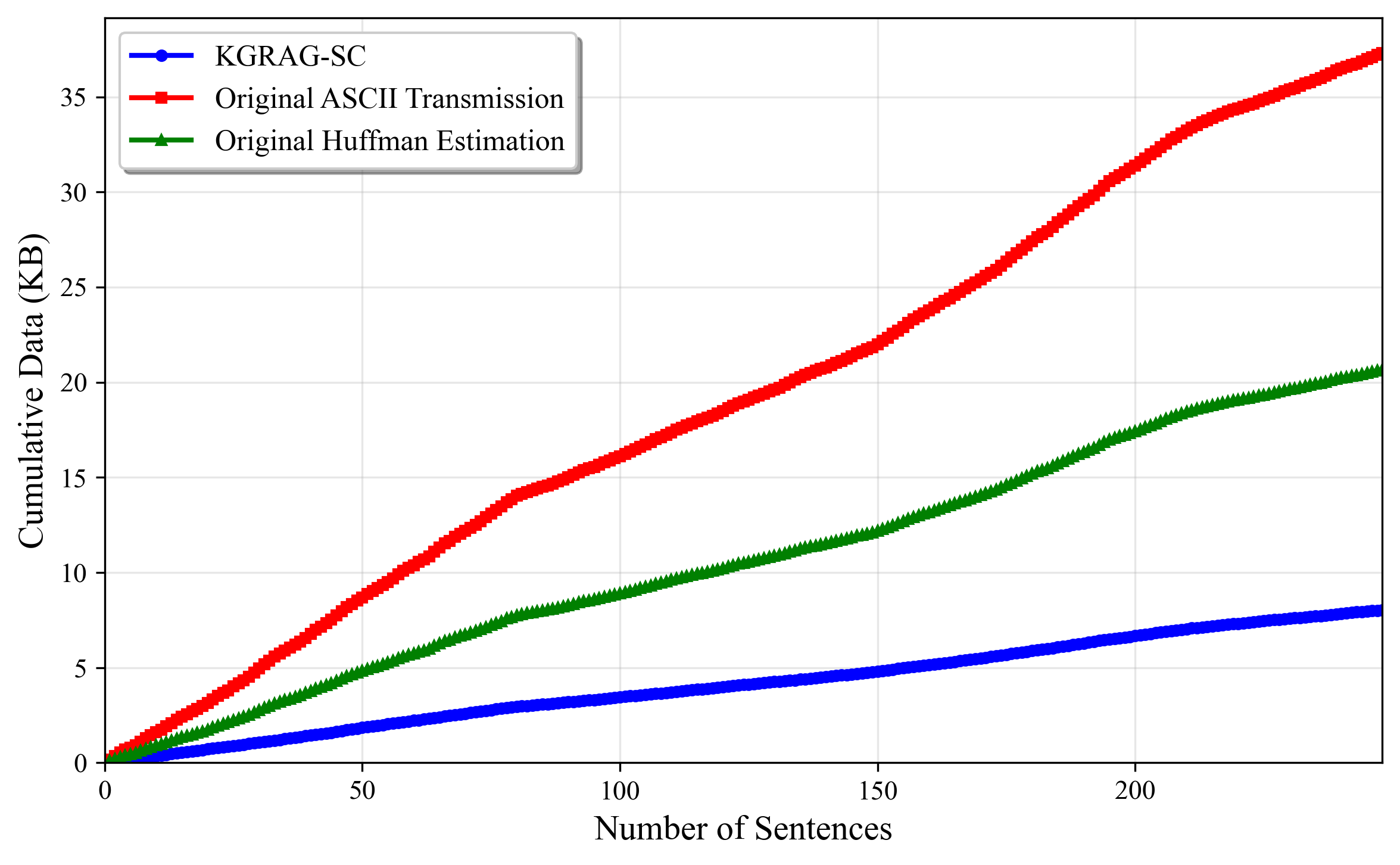}
\caption{Comparison of cumulative transmitted data volume over a sequence of 250 sentences.}
\label{fig:cumulative_comp}
\end{figure}

The efficiency of KGRAG-SC is evaluated in Fig.~\ref{fig:bits_comp} and Fig.~\ref{fig:cumulative_comp}. These results clearly demonstrate the significant reduction in data overhead achieved by transmitting a compact set of entity IDs instead of the full source text.

Fig.~\ref{fig:bits_comp} illustrates the number of bits required to transmit each sentence in our test corpus. KGRAG-SC consistently requires a significantly lower and more stable number of bits per sentence compared to both raw ASCII and Huffman-coded text. This highlights a key advantage of KGRAG-SC: the transmission cost is coupled with the underlying semantic complexity of the sentence, rather than its length or verbosity.

The long-term benefits of this efficiency are further emphasized in Fig.~\ref{fig:cumulative_comp}, which plots the cumulative data volume transmitted over 250 sentences. The performance gap between KGRAG-SC and the baseline schemes widens progressively, showcasing substantial bandwidth savings over time. By distilling each sentence down to its core semantic entities, KGRAG-SC achieves a far more parsimonious use of communication resources, which is a critical advantage for bandwidth-constrained applications.

\section{CONCLUSION AND FUTURE WORK}

In this paper, we proposed and evaluated KGRAG-SC, a novel SC scheme that integrates a knowledge graph with a large language model. KGRAG-SC demonstrates significant improvements in both semantic fidelity, especially under low SNR conditions, and transmission efficiency compared to traditional communication methods. By transmitting a structured, condensed semantic representation instead of the raw text, KGRAG-SC proves highly robust to channel noise.

Building on this work, several promising directions for future research emerge:

\textbf{Dynamic Knowledge Graph Management:} The current schemes relies on a static, pre-shared knowledge graph. A key area for future work is to develop mechanisms for dynamic KG updates, including knowledge addition, revision, and forgetting. This would enable the communication system to adapt to evolving contexts and new information in real-time, making it suitable for more dynamic and non-stationary environments.

\textbf{Privacy-Enhanced SC:} To address the transmission of sensitive information, future iterations could incorporate a privacy attribute for entities within the knowledge graph. This attribute could trigger specialized processing pipelines, such as applying differential privacy techniques, selective encryption, or transmitting only generalized, less sensitive information for designated private entities, thus enabling secure and context-aware communication.

\textbf{End-to-End System Optimization:} While the current components are highly effective, the overall system could be further enhanced through end-to-end optimization. This would involve jointly training the semantic extraction, importance measurement, and text reconstruction modules. Such an approach could allow the system to learn optimal transmission strategies automatically, dynamically balancing compression rate, semantic accuracy, and channel robustness based on the specific content and channel state.

\bibliographystyle{IEEEtran}
\bibliography{IEEEabrv,bib}

\end{document}